# Solid phase epitaxy of ferromagnetic MnAs dots on GaMnAs layers

*Janusz Sadowski (1), Martin Adell (2), Janusz Kanski (2), Lars Ilver (2), Elzbieta Janik (1),*

*Elzbieta Lusakowska (1), Jaroslaw Z. Domagala (1), Rimantas Brucas (2), Maj Hanson (2)*

(1) Institute of Physics, Polish Academy of Sciences, al Lotników 32/46, 02-668 Warsaw, Poland

(2) Department of Experimental Physics, Chalmers University of Technology, SE-41296 Göteborg, Sweden

Abstract:

Formation of MnAs quantum dots in a regular ring-like distribution has been found on MBE-grown (GaMn)As surfaces after low-temperature annealing under As capping. The Mn was supplied by out-diffusing Mn interstitials from (GaMn)As. With 5 at% substitutional Mn the quantum dots appeared for (GaMn)As layers thicker than 500 Å. For thinner layers the Mn-rich surfaces, presumably monolayer thick MnAs, are smooth and well-ordered (1x2), and are well suited for continued epitaxial growth.

---

* current address: Institut für Angewandte und Experimentelle Physik, Universität Regensburg, 93040 Regensburg, Germany; phone: +49 9419432064, e-mail: janusz.sadowski@physik.uni-regensburg.de

In the quest for magnetic semiconductors for spintronics applications GaMnAs remains the best candidate with the carrier induced ferromagnetism and relatively high ferromagnetic phase transition temperature (Tc)[1, 2.] However the potential applications of GaMnAs need Tc to be higher than the present limit of 160 – 170 K[3 – 6]. One way to get around this problem is to use hybrid ferromagnetic metal/semiconductor structures, which can provide spin-polarized carriers for injection into a semiconducting material. With this type of structures, extensively investigated by several research groups[7-10], the efficiency of spin-injection is, however, very limited. One complication is the usually disordered interface between the semiconductor and the ferromagnetic metal, resulting in most cases from the incompatibility of the different materials. In this letter we report on a new method for obtaining thin epitaxial very Mn rich (presumably MnAs) epilayers on GaMnAs by making use of a specific property of GaMnAs, namely the presence of highly mobile Mn interstitials ($Mn_I$) within the bulk of MBE grown GaMnAs layers[11–14]. We also show that by the same process we can get self-organized ring patterns of MnAs quantum dots on the (GaMn)As surface. Another aspect of this letter is the identification of an efficient mechanism for removal of $Mn_I$ from GaMnAs.

After the studies of Masek et. al.[15] and Yu et. al.[12], Mn interstitials are recognized to be the prevalent defects in (GaMn)As, degrading its magnetic and transport properties[14,18]. Efficient removal of $Mn_I$, which is achieved by the low temperature post-growth annealing, is thus the key for obtaining high quality ferromagnetic (GaMn)As based structures. From the different results of annealing studies of oxidized[6] and LT GaAs capped[16,17] GaMnAs, it has become clear that the state of (GaMn)As surface is crucial. Indeed, by carrying out the annealing under As capping, we were recently able to demonstrate that annealing process is dramatically accelerated, which shows that surface trapping of the diffusing $Mn_I$ is the key ingredient in the annealing process[19]. As we show here, the reaction between the out-diffusing Mn and the As capping can result in either a continuous and smooth and well-ordered surface, or in a new kind self-organized MnAs quantum dots pattern.

The GaMnAs samples were grown in a KRYOVAK MBE system, located at one of the beamlines at the MAX-lab storage ring. The MBE growth was started by a standard high temperature (HT) GaAs buffer, grown at a substrate temperature of 600 $^o$C. Then the substrate temperature was decreased to 240 $^o$C and the (GaMn)As layer was deposited. The Mn content in the present samples was 5%, calibrated by means of reflection high energy electron diffraction (RHEED) intensity oscillations. The RHEED oscillations were recorded during growth of all samples, which means that the layer thickness was controlled to one molecular layer (ML) precision (2.84 Å in the case of $Ga_{0.95}Mn_{0.05}As$). Directly after the MBE growth the (GaMn)As samples were cooled to below 100 °C, and the surface was exposed to the $As_2$ flux from an As cracker effusion cell for 30 – 60 min to form an amorphous As capping layer, about 2000 Å thick. Unlike the situation with the bare (GaMn)As[4-6], post-growth annealing of the As-capped surface can be performed in vacuum or in air with the same efficiency, as the reacting interface is protected.

In this report we discuss the effects of annealing on the surface morphologies of $Ga_{0.95}Mn_{0.05}As$ samples with the thicknesses of 1.0, 0.3, 0.2, 0.15, and 0.1μm. Upon deposition of the As layer the RHEED picture changed quickly (5 – 10 s after opening As shutter) from a (1x2) reconstructed, streaky image characteristic for smooth 2D (GaMn)As surface, to a diffuse intensity distribution characteristic for an amorphous surface. After the As deposition each sample was taken out of the vacuum system and cleaved into two pieces. One piece was reintroduced to the MBE system, and the As capping was desorbed. During this process the state of the (GaMn)As surface was monitored by RHEED. To achieve the As capping desorption, the sample was annealed first at a temperature of 180 $^o$C for 3h (this was found before as an optimum treatment for removing practically all the $Mn_I$ from the (GaMn)As bulk[19]), then the sample temperature was raised quickly to 270 – 290 $^o$C in order to completely desorb the remaining capping layer. The As capping desorption was recognized by the disappearance of the diffused background in the RHEED image, and appearance of the clear RHEED diffraction patterns. Depending on the (GaMn)As layer thickness, the RHEED image showed either purely 2 dimensional (2D) streaky diffraction pattern resulting from 2D diffraction on smooth, monocrystalline surface for GaMnAs samples with thickness equal to 1000 Å and below – see Figure 1,a and 1b; or a mixture of 2D and 3D spotty image for thicker samples - see Fig.1.c and 1d. The diffraction spots in Fig. 1.d indicate the presence of 3D objects on the surface, which were identified as quantum dots, as show in Fig.1c. The elemental composition of the dots is not determined, but considering the preparation method and the RHEED patterns, we are convinced that they consist of hexagonal MnAs. The AFM image shows that the distribution of MnAs dots is not random – they are preferentially located on a ring-like structure.

This tendency for an ordered distribution of the MnAs dots can also be seen in the case of samples 2-4. Fig. 2 shows the AFM images of all samples. For samples 1-3 the density of the MnAs dots tends to increase with increasing layer thickness. This trend can be understood simply as an effect of increasing amount of out-diffused Mn, since the amount of interstitial Mn in the (GaMn)As should be directly proportional to the layer thickness, and the out-diffusion appears to be equally efficient for all thicknesses. It is interesting to note that the dimensions of individual MnAs dots are not dependent on the (GaMn)As thickness. The reason for a deviating behavior for the thinnest sample is unclear. At this point we are also unable to provide an explanation for the spontaneous ordering of the MnAs dots, though on may speculate that it may be connected with the microscopical mechanism of diffusion of $Mn_I$ from the (GaMn)As bulk. This process remains to be explained.

We have also investigated how the properties of GaMnAs layers beneath the MnAs dots, changed due to the annealing process. It was already observed, that the out-diffusion of $Mn_I$ defects is associated with the significant reduction of the GaMnAs lattice parameter[20,21]. Fig.3 shows the results of X-ray diffraction (XRD) measurements of sample 4. The omega scans for 004 Bragg reflections of non annealed and annealed pieces are shown. The significant shift of GaMnAs (004) Bragg diffraction peak towards higher angles prove the annealing induced reduction of GaMnAs lattice parameter ($a_{GaMnAs}$). The changes of $a_{GaMnAs}$ are caused by removing Mn from interstitial sites[19-22].

Since bulk MnAs is ferromagnetic at room temperature, we have also applied Magnetic Force Microscopy (MFM) to study the quantum dot decorated surfaces. The magnetic imaging of the MnAs/GaMnAs structures was performed with a Digital Instruments Dimension 3000 Scanning Probe

Microscope (SPM) operated in tapping-lift mode. This mode allows simultaneous acquisition and clear separation between the topographic and the magnetic data. Standard commercially available tips with CoCr coating were used for the image acquisition. All the tips were magnetized vertically with a permanent magnet (along the needle). The MFM image contrast is proportional to the gradient of the magnetic force between tip and sample. In order to minimize the tip influence on the magnetic structure, to verify the reproducibility and to exclude artifacts, a set of measurements was performed on the same sample at different scan angles and the tip-to-sample distance was changed in the range from 30 nm to 150 nm during magnetic imaging. The observed magnetic contrast was changing with the tip-to-sample distance but the magnetic features of the sample were essentially the same. The images were analyzed by applying a two-dimensional Fourier transform (2DFT) algorithm. The 2DFT analysis enabled determination of the average size and distribution of the particles. The results are presented in the Fig.4. Comparing AFM and MFM images, for the same sample area (Fig.4a, and 4b, respectively) is evident that the largest MnAs dots give some magnetic contrast, whereas smaller dots are not visible. This signifies that either the small dots are not ferromagnetic, or they are ferromagnetic at temperatures lower than the MFM measurement temperature.

In summary – we have demonstrated a new method for obtaining MnAs quantum dots using solid phase reaction between Mn out-diffused from interstitial sites in (GaMn)As layers and an adsorbed As capping layer. The density of the MnAs dots depends on the thickness of the initial (GaMn)As layer, as expected. An unexpected (and unexplained) ordering of the MnAs dots is observed, with the particles spontaneously forming a ring-type pattern. The feasibility of solid phase epitaxy of MnAs (thin layers and dots) on (GaMn)As opens new possibilities of obtaining composite structures for development of spintronics materials.

Acknowledgements

This work was partially supported by the Swedish Research Council (VR). One of the authors (J.S) acknowledges the financial support from the Polish State Committee for Scientific Research (KBN) through the grant PBZ-KBN-044/P03/2001.

Figure captions

Fig.1. Surface morphologies of GaMnAs 1000 Å, and 1 μm thick samples after desorption of arsenic capping layer at high vacuum conditions.

(a), (b) AFM and RHEED diffraction images from 1000 Å thick sample with smooth surface after As desorption );

(c), (d) GaMnAs surface with MnAs dots seen as islands in AFM and 3D diffraction features in RHEED image

Fig.2. AFM detected surface morphologies of MnAs dots, on GaMnAs layers with different thicknesses: (a) - 1 μm, sample 1; (b) - 0.3 μm, sample 2; (c) - 0.2 μm, sample 3; (d) - 0.15 μm, sample 4

Fig. 3. X-ray (004) Bragg reflection from 0.15 μm thick GaMnAs layer (sample 4) before annealing (solid curve) and after annealing at high vacuum (dashed curve) leading to the out diffusion of Mn interstitials from GaMnAs bulk and formation of MnAs dost on the surface. Out diffusion of Mn interstitials reduces the GaMnAs lattice parameter, evidenced by the significant shift of (004) GaMnAs reflection towards higher diffraction angles.

Fig. 4. (a) - AFM, and (b) - MFM images from the same surface region of annealed 0.2 μm thick GaMnAs layer with MnAs dots on the surface.

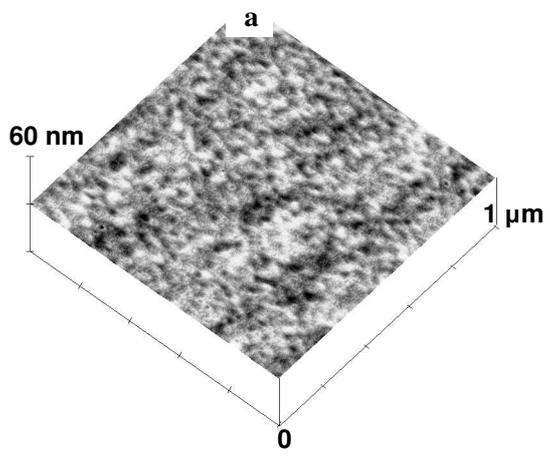
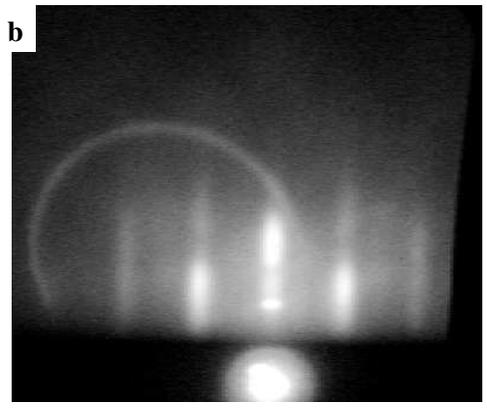
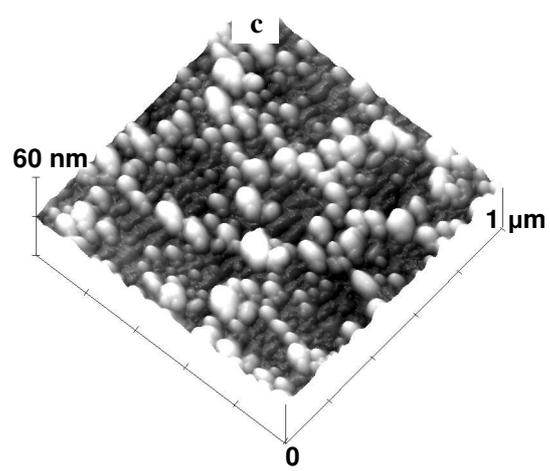
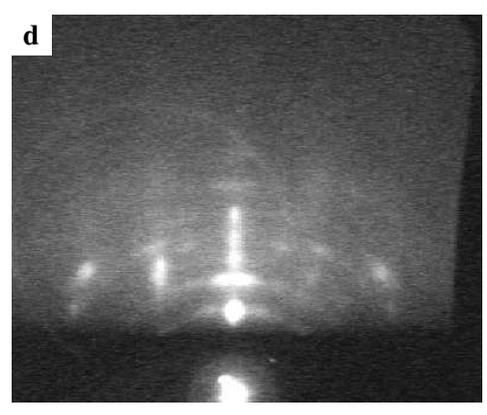

*J. Sadowski et al.  Fig. 1*

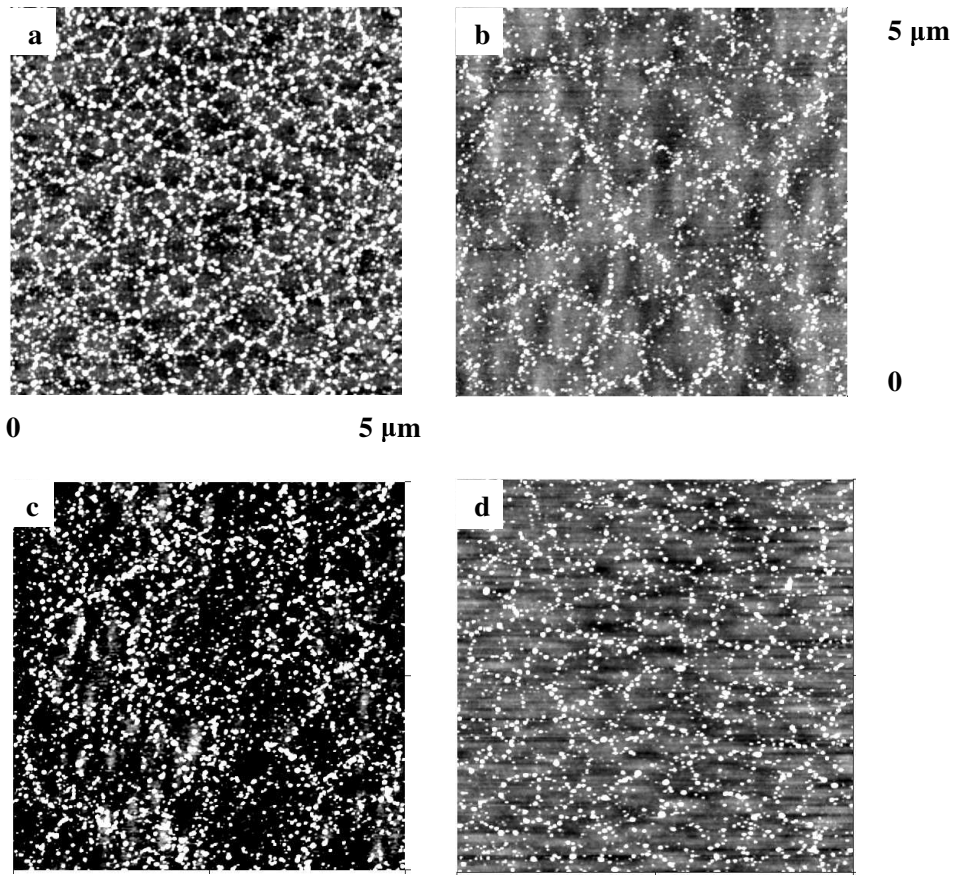

*J. Sadowski et al.  Fig. 2*

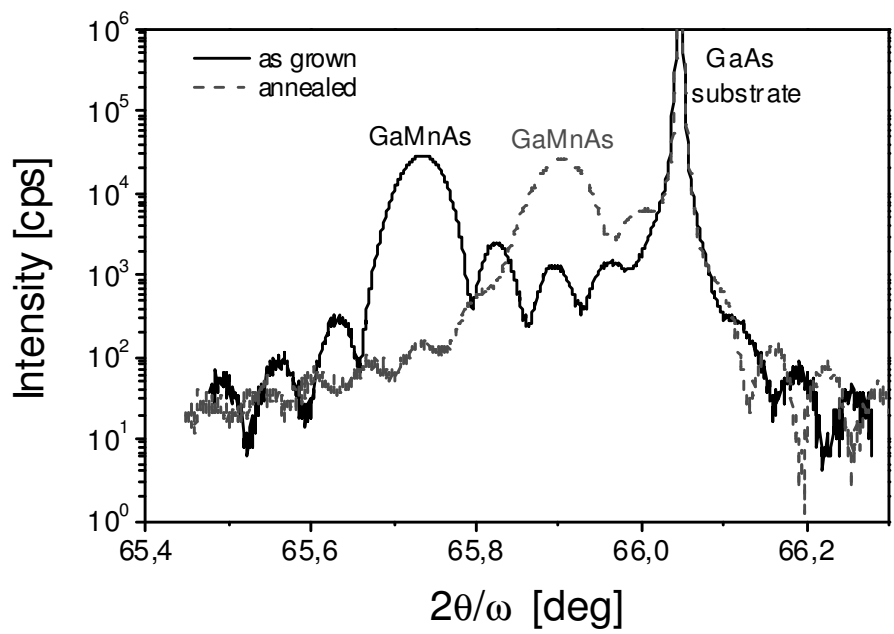

*J. Sadowski et al. Fig. 3*

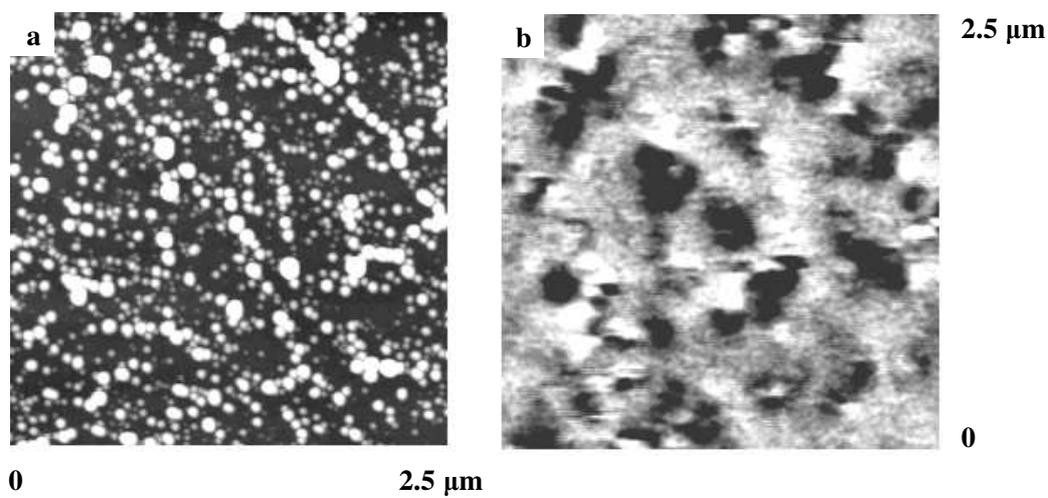